# Test-retest Reliability of Psychophysical Tasks Using Structured Light-Induced Entoptic Phenomena

Taranjit Singh[1,2], Mukhit Kulmaganbetov[1,2,3], Zhangting Wang[1], Dmitry Pushin [1,4,5,6], Benjamin Thompson[1,2,7], Andrew Silva[8], Melanie Mungalsingh[7], Iman Salehi[7], Davis Garrad[5], David Cory[4,9], Dusan Sarenac[1,6,7,10]

[1] Centre for Eye and Vision Research, 17W Hong Kong Science Park, Hong Kong
[2] Entoptica Limited, 136 Bonham Strand, Sheung Wan, Hong Kong
[3] Kazakh Eye Research Institute, Almaty, Kazakhstan, A05H2A8
[4] Institute for Quantum Computing, University of Waterloo, Waterloo, ON, Canada, N2L3G1
[5] Department of Physics and Astronomy, University of Waterloo, Waterloo, ON, Canada, N2L3G1
[6] Incoherent Vision Inc., Wellesley, ON, Canada, N0B2T0
[7] School of Optometry and Vision Science, University of Waterloo, Waterloo, ON, Canada, N2L3G1
[8] Department of Psychology, Idaho State University, Pocatello, Idaho 83209, USA
[9] Department of Chemistry, University of Waterloo, Waterloo, ON, Canada, N2L3G1
[10] Department of Physics, University at Buffalo, State University of New York, Buffalo, New York 14260, USA

**Abstract**

Structured light (SL)-induced polarization perception presents a promising method for the early detection of macular diseases such as age-related macular degeneration. We investigated the test-retest reliability of a psychophysical task using SL-based stimuli to induce entoptic patterns in individuals with healthy vision. Twenty-eight participants underwent thorough eye examinations to confirm they had healthy eyes and good vision (logMAR BCVA ~0.00). Of these, 25 participants (n=50 eyes) aged 21 to 75 completed two identical tasks separated by 1 to 14 days. Using SL-based stimuli that produced a rotating entoptic pattern containing 22 azimuthal brushes, we measured the retinal eccentricity threshold $R_T$ at which participants could reliably identify the direction of rotation by varying the size of a central obstruction. This threshold reflects the visual angle of the pattern for each participant. We calculated the reliability coefficient (intraclass correlation coefficient, ICC) using a two-way mixed-effects model and conducted a Bland-Altman analysis to assess test-retest reliability. The ICC was 0.83 [95% CI: 0.62 - 0.93] for right eyes (RE) and 0.93 [95% CI: 0.84 - 0.97] for left eyes (LE), indicating good reliability. The Bland-Altman analysis showed a mean difference of -0.32° (SD: 1.51) and -0.12° (SD: 1.03) for RE and LE respectively between the first and second sessions, with limits of agreement ranging from -3.28° to 2.64° and -2.14° to 1.89° (Figure 1) for RE and LE respectively, confirming strong agreement and no significant bias. These results demonstrate that SL-based psychophysical tasks are a reliable method for assessing polarization perception, potentially improving screening for diseases affecting macular health.

**Introduction**

The assessment of visual perception, particularly in relation to macular health, has become an emerging focus within vision science [1,2]. Human perception of polarized light is enabled through the Haidinger's brushes phenomenon [3] though it deteriorates with the progression of macular disorders [4-7]. Unfortunately, the intrinsically faintness, transience, and low contrast of the Haidinger's brushes percept have constrained its clinical adoption, as it remains difficult to perceive reliably even in individuals with healthy vision. [5-7].

Structured light (SL), characterized by tailored spatial and polarization profiles, enables the generation of novel stimuli that enhance polarization-dependent entoptic phenomena through controlled azimuthal modulation [8]. SL stimuli are emerging as a promising new method for psychophysically assessing macular health [8-14] because they provide a larger and more readily detectable entoptic percepts relative to traditional polarization-based entoptic methods [13]. For example, SL techniques can serve as measures of circularly oriented macular pigment optical density (coMPOD) [14]. This measurement may offer a functional basis for detecting early disruptions in the macula associated with diseases like age-related macular degeneration (AMD), a prospect highlighted for future clinical translation [16-18].

Conventional methods for assessing visual perception, such as perimetry and various visual field tests, exhibit varying levels of repeatability, influenced by factors such as individual differences in attention, cognitive load, and environmental conditions [22-24]. This variability can lead to inconsistent outcomes, raising concerns about the reliability of these assessments in both clinical and research settings [22]. Psychophysical methods such as macular pigment optical density (MPOD) measurements via heterochromatic flicker photometry, have also faced challenges regarding reliability and consistency [24-27]. Additionally, traditional methods that rely on Haidinger's brushes also encounter similar issues [3-6].

In this study, we aimed to evaluate the test-retest reliability of an SL-based psychophysical task that utilizes SL-induced entoptic phenomena [8-15]. While the results of SL-based perception tests are influenced by macular health, they are also subjective, introducing a degree of individual variation [8,9,13-15,22,28]. Through this investigation, we aim to contribute to the growing body of knowledge surrounding structured light applications in vision science, paving the way for enhanced diagnostic tools that leverage the unique properties of polarized light perception.

**Methods**

*Screening*

A total of 28 participants were recruited from the Centre for Eye and Vision Research (CEVR). Every participant was subjected to inclusion screening where they underwent an eye and vision examination including assessment of medical and family history, unaided visual acuity, subjective refraction using trial frames and lenses, binocular vision tests, ocular motility, contrast sensitivity using the Thomson Chart (Thomson Software Solutions, UK), slit-lamp bio-microscopy, indirect ophthalmoscopy, colour fundus photography (Nidek AFC-330, Japan), optical coherence tomography (Topcon DRI OCT Triton, Japan), ocular biometry (Zeiss IOLMaster 700, Germany), and macular pigment optical densitometry (MPS II, UK). 25 participants with good vision (best-corrected visual acuity of 0.00 logMAR or better in each eye) and healthy eyes (with no history of eye diseases, ocular injuries, neurological disorders, etc.) were recruited, 3 were eliminated or withdrew from the study.

*Psychophysical Task*

The psychophysical task conducted in this study utilized structured light (SL)-based stimuli [10,13-15] that produced entoptic patterns composed of $N_f$=22 brushes for a duration of 0.5 seconds per trial. Participants were instructed to identify the direction of rotation of the entoptic patterns, which could be either clockwise or counter-clockwise, as dictated by a motorized polarizer [8,13-15].

To assess retinal eccentricity of their SL perception, we employed structured light techniques to block central areas of varying sizes in the generated entoptic patterns. The threshold mask was created by a spatial light modulator (SLM). The radius of the obstruction was adjusted according to the 2-up/1-down staircase method that has been described previously [14]. The staircase commenced with a central mask radius of 0.45° visual angle and changed with a step size of 1.35° visual angle in radius for the first three reversals, then with a step size of 0.90° for the next three reversals, then became 0.45° for the next three reversals and finally changed to 0.23° for the remaining reversals. This method enables a performance accuracy measurement of 70.7% for the threshold radius of each stimulus [14] and retinal eccentricity $R_T$ of the perceivable area of the SL-based stimuli.

Note that to convert the radius of the central mask from SLM pixel units to visual angle degrees, a global fixed conversion ratio of 0.046 degrees per pixel, as established in a previous study [15] to be a reliable average, was applied. This conversion rescales the data into interpretable degrees of visual angle without altering the shape or distribution of the data.

$R_T$ estimates the visual angle of the observed entoptic pattern, which becomes visible due to the arrangement of radially oriented birefringent layers of photoreceptor axons and the dichroism of macular pigments located in Henle's fiber layer of the retina, which filters blue azimuthally polarized light [13,15].

Participants were required to remove any refractive corrections, including contact lenses, prior to the experiment, which was conducted monocularly. Both eyes were tested on the same day, with a brief interval of 1 to 2 minutes allowed for chinrest adjustments after the first eye was tested. The first eye examined was chosen randomly, with the other eye covered; the second eye was then tested under identical conditions.

An identical test was conducted for each participant on a separate day, with the interval between the two tests ranging from 1 to 14 days. The first eye examined was selected randomly, followed by testing the other eye under the same conditions. In total, two identical psychophysical tasks were administered to 25 participants (n=50 eyes) aged 21 to 75 years.

Both participants and experimenters were aware that the second session was a repeat of the first (i.e., the retest was not conducted under blind conditions). This was unavoidable because the psychophysical procedure requires explicit subject instruction and feedback, and the technical setup was identical across sessions. However, the objective nature of the response (forced-choice discrimination of stimulus rotation) and the automated staircase procedure minimize experimenter influence.

*Statistical analysis*

Two values of $R_T$ were obtained from the two identical psychophysical tasks for each eye, right eye (RE) and left eye (LE). Interclass correlation coefficients (two-way mixed- effects model) and Bland Altman plots were used to assess test-retest reliability. In addition, the Pearson correlation coefficient between the $R_T$ value in the first and second session were calculated to assess correlation of results between the two sessions. Data from left and right eyes were analysed separately [32].

**Results**

Table 1.  Mean $R_T$ of all participants; right eye and left eye

|  | Right Eye (n=25) | | Left Eye (n=25) | |
| --- | --- | --- | --- | --- |
| Session | 1st | 2nd | 1st | 2nd |
| Mean $R_T$ ± SD | 5.48° ± 0.20° | 5.79° ± 0.19° | 5.94° ± 0.16° | 6.06° ± 0.18° |

The interpretation of ICC values is critical for determining the quality of the reliability of measurements, especially for measurements from a response-based task. As a general guideline, ICC values above 0.75 suggest good to excellent reliability, indicating that the measurements are highly reliable and can be used confidently in clinical or research settings [22,28-30]. The test-retest reliability was calculated as the intraclass correlation coefficient (ICC) using a two-way mixed-effects, single-rater model (ICC(3,1)).[1]

ICCs were 0.83 [95% CI: 0.62 - 0.93] for RE and 0.93 [95% CI: 0.84 - 0.97] for LE, indicating good to excellent test-retest reliability for each eye [22,28-30].

The Bland-Altman analysis revealed a non-significant mean difference of -0.32° (SD: 1.51) and -0.12° (SD: 1.03) for RE and LE respectively between the first and second sessions, with limits of agreement ranging from -3.28° to 2.64° and -2.14° to 1.89° (Figure 1) for RE and LE

---

[1] The calculation was based on the following formulation, which quantifies the correlation between two test sessions:

$$r = \frac{1}{Ns^2} \sum_{n=1}^{N} (x_{n,1} - \bar{x})(x_{n,2} - \bar{x})$$

where $N$ is the number of subjects, $s^2$ is the pooled variance across both sessions, $x_{n,1}$ and $x_{n,2}$ are the measurements for the $n$-th subject in session 1 and session 2, respectively, and $\bar{x}$ is the overall mean across all measurements from both sessions.

respectively. In addition, the Pearson correlation coefficient between the $R_T$ value in the first and second session was 0.71 for RE and 0.87 for LE respectively with a p-value < 0.01 for both eyes (Figure 2) indicating good correlation of results between the two sessions.

Table 2. Two-way mixed-effects ICC values of individual eyes

|  | Right Eye | Left Eye |
| --- | --- | --- |
| ICC [95% CI] | 0.83 [0.62, 0.93] | 0.93 [0.84, 0.97] |

Table 3. Bland Altman Analysis of individual eyes

|  | Right Eye | Left Eye |
| --- | --- | --- |
| Mean Difference (± SD) | -0.32° ± 1.51° | -0.12° ± 1.03° |
| Limits of Agreement | -3.28° to 2.64° | -2.14° to 1.89° |

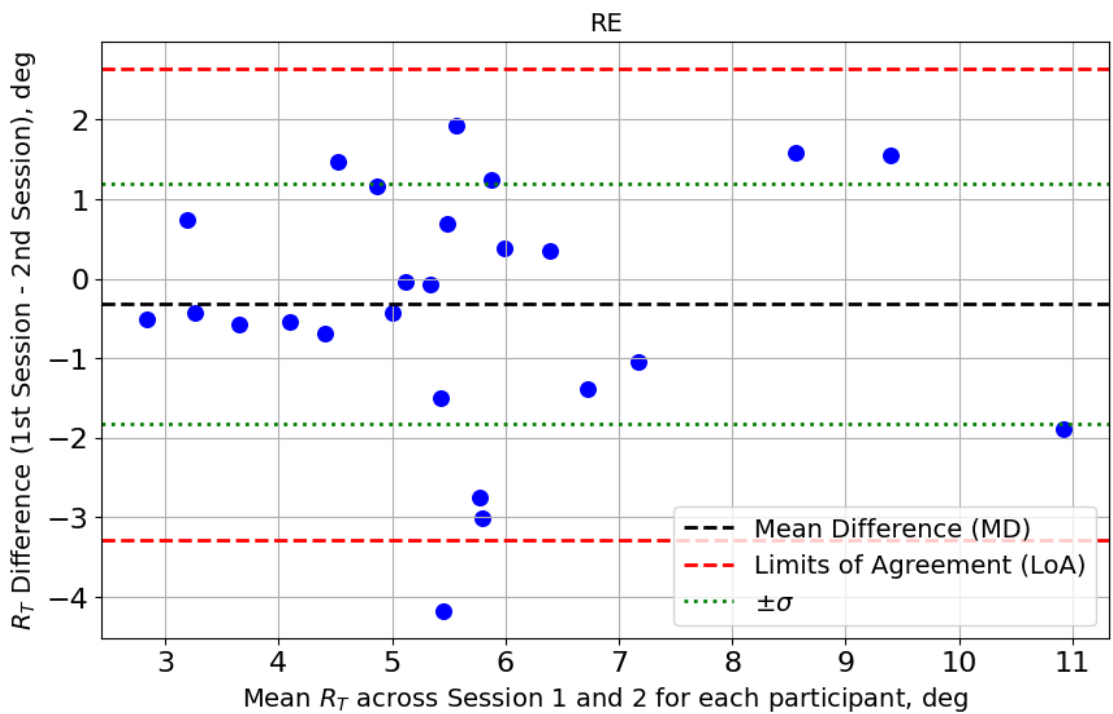

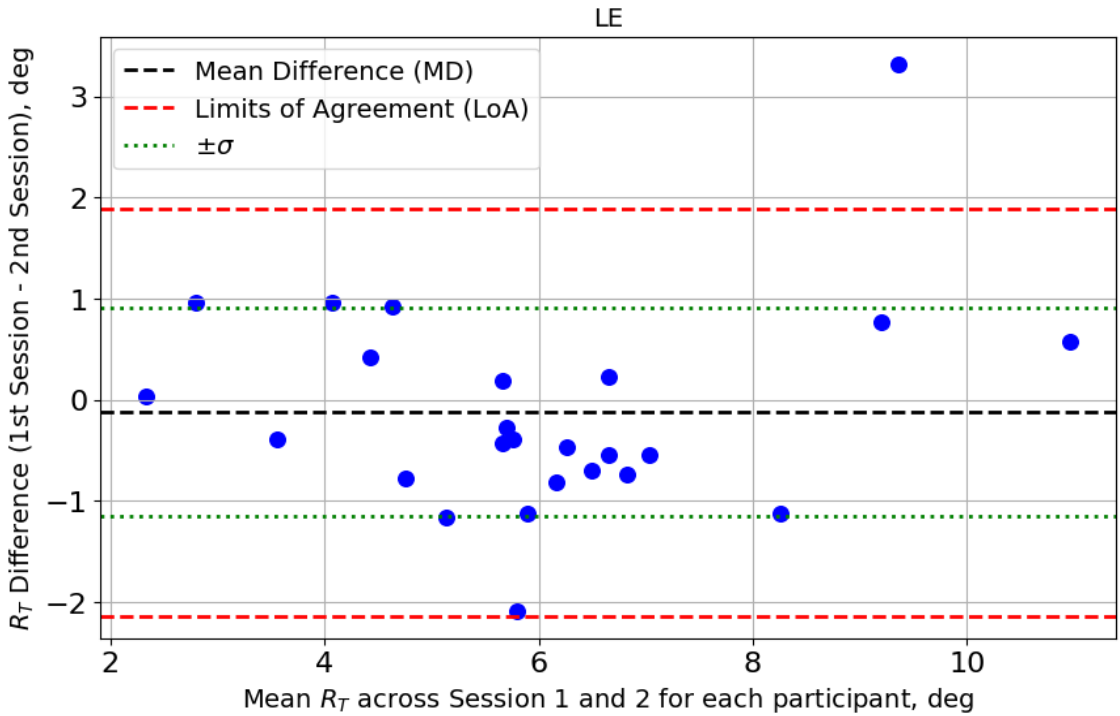

Figure 1. Bland-Altman plot with mean $R_T$ across Session 1 and 2 against the difference of $R_T$ over the two sessions for each participant, for RE (top) and LE (bottom). There was nearly zero bias between the 1st and 2nd Session for both eyes.

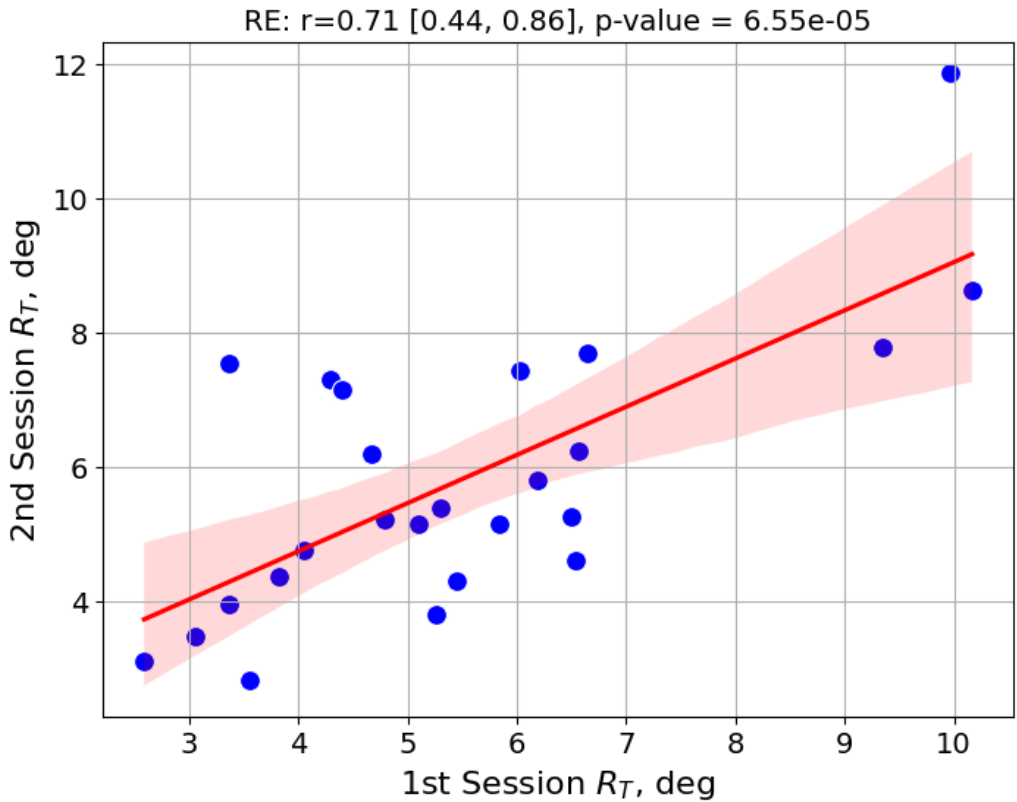

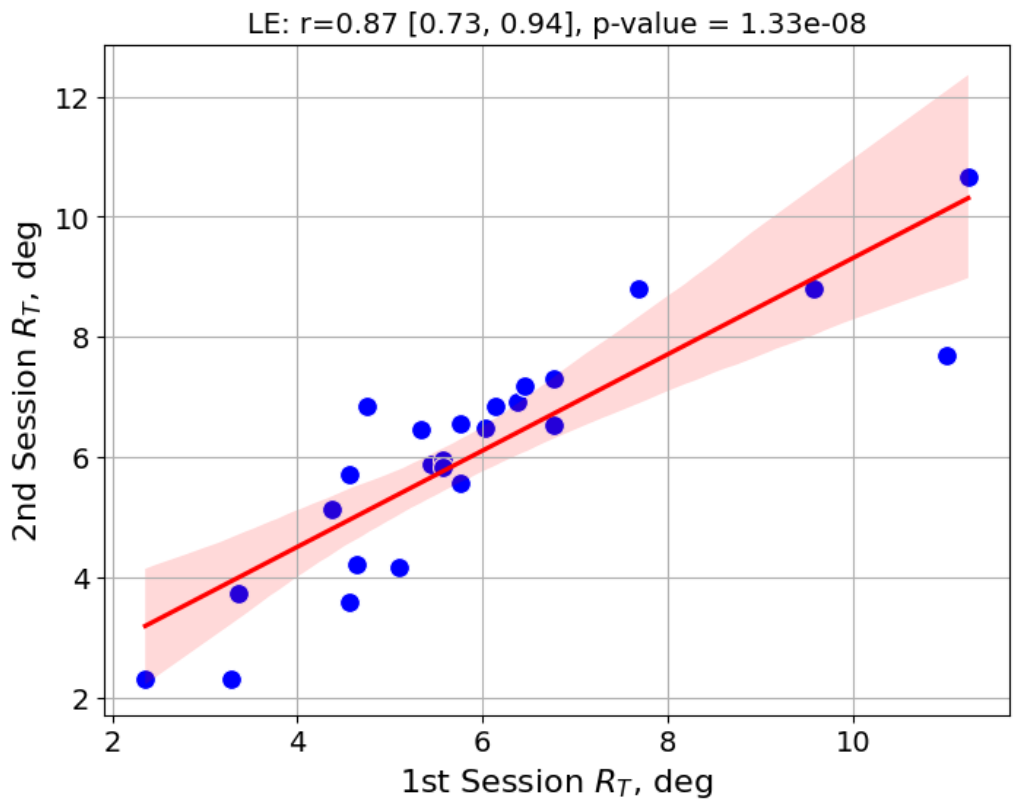

*Figure 2. Linear correlation plot between $R_T$ values of first and second session. It shows a Pearson correlation coefficient (r), confidence interval and p-value of $R_T$ for RE (top) and LE (bottom); indicating a strong association of results for both eye between the two sessions.*

**Discussion**

When interpreting the Intraclass Correlation Coefficient (ICC) values for test-retest reliability, it is essential to recognize that different studies and fields may adopt varying guidelines. While no universally accepted standard exists, several common interpretations have emerged in the literature. For instance, Koo and Li (2016) propose that ICC values below 0.50 indicate poor reliability, those between 0.50 and 0.75 suggest moderate reliability, values from 0.75 to 0.90 indicate good reliability, and values above 0.90 signify excellent reliability [31]. In contrast, Cicchetti (1994) presents a slightly different framework, categorizing values under 0.40 as poor, between 0.40 and 0.59 as fair, from 0.60 to 0.74 as good, and from 0.75 to 1.00 as excellent [30]. These varying interpretations underscore the necessity of careful consideration of context and application when evaluating reliability based on ICC values [33].

Building on this foundation, an average ICC value of 0.88 (RE=0.83, LE=0.93) in our study can be classified as reflecting good to excellent reliability. This finding indicates that the SL-based psychophysical method is a reliable approach for testing polarization perception in the human eye.

However, it is important to address potential limitations that may arise from this method. Specifically, the first test, even when administered identically, could influence the results of subsequent tests. Given the psychophysical nature of the task, there is a possibility of a natural learning effect that may enhance participants' performance during the second test administration, potentially resulting in a higher $R_T$ value.

One notable study employed a similar psychophysical task and investigated the effects of eye dominance and testing order on the resulting thresholds [15]. A random eye was selected to perform the psychophysical task first, followed immediately by the other eye and their analysis revealed no statistically significant difference between the first and second eye tested [15]. In the current study, the mean difference between the first and second sessions was minimal, at -0.22°, with approximately 40% of participants demonstrating a lower $R_T$ value in the second session compared to the first. These findings suggest that learning did not influence the outcomes of this psychophysical task and its test-retest reliability [15,22].

A limitation of this study is that all participants had normal vision. Further work is required to assess whether individuals with early-stage macular disease exhibit similar test-retest reliability.

**Conclusions**

SL-based psychophysical tasks are a reliable method for assessing polarization perception. By providing a quantitative measure of entoptic patterns, this technique has the potential to enhance screening for diseases affecting macular health.

**Acknowledgements**

This study was funded in part by the InnoHK initiative of the Innovation and Technology Commission of the Hong Kong Special Administrative Region Government. This research was also undertaken thanks in part to funding from the Canada First Research Excellence Fund (CFREF), the New Frontiers in Research Fund (NFRF), the Canadian Excellence Research Chairs (CERC) program and the Natural Sciences and Engineering Research Council of Canada (NSERC).